\documentclass[onecolumn,floatfix,superscriptaddress,showpacs,showkeys,nofootinbib,preprint]{revtex4}

\usepackage{epsfig}
\usepackage{amssymb,latexsym,amsmath}
\newcommand{\eq}[1]{\begin{align} #1 \end{align}}

\begin{document}

\title{Transparency, Mixing and Reflection
of  Initial Flows \\
in Relativistic Nuclear Collisions
}

\author{Marek Ga\'zdzicki}
\affiliation{Institut f\"ur Kernphysik, Johann Wolfgang
 Goethe Universit\"at Frankfurt, Germany}
\affiliation{\'Swi\c{e}tokrzyska Academy, Kielce, Poland}
\author{ Mark Gorenstein}
\affiliation{Bogolyubov Institute for Theoretical Physics, Kiev, Ukraine}
\affiliation{Frankfurt Institute for Advanced Studies, Frankfurt, Germany}

\begin{abstract}
We propose to use the hadron number fluctuations in the limited
momentum regions to study the evolution of initial  flows in high
energy nuclear collisions. In this method 
by a proper preparation of a collision sample
the projectile and target
initial flows are marked in fluctuations  in the number of colliding
nucleons. We discuss
three limiting cases of the evolution of flows, transparency, mixing
and reflection, and present for them quantitative predictions
obtained within several models. Finally, we apply the method to the
NA49 results on fluctuations of the negatively charged hadron
multiplicity in Pb+Pb interactions at 158$A$ GeV and conclude that
the data favor a hydrodynamical model with a significant degree of
mixing of the initial flows at the early stage of collisions.



\end{abstract}
\maketitle

{\bf 1.} The main goal of investigations of high energy
nucleus-nucleus (A+A) collisions is to uncover properties of
strongly interacting matter at high energy densities and, in
particular, to look for its hypothetical phases and transitions
between them. Qualitative features of the rich experimental data
collected thus far indicate that the produced matter experiences
strong collective expansion and it is close to local equilibrium
\cite{qm2004}. Moreover, the properties of the matter change rapidly
at the low CERN SPS energies ($\sqrt{s_{NN}} \approx$ 8 GeV)
suggesting the onset of deconfinement and thus the existence of a
new state of matter, a Quark Gluon Plasma~\cite{na49anomalies,smes}.
The properties of this new phase are under active studies in A+A
collisions at the BNL RHIC \cite{rhic} ($\sqrt{s_{NN}} =$ 200 GeV).

\begin{figure}[ht!]
\epsfig{file=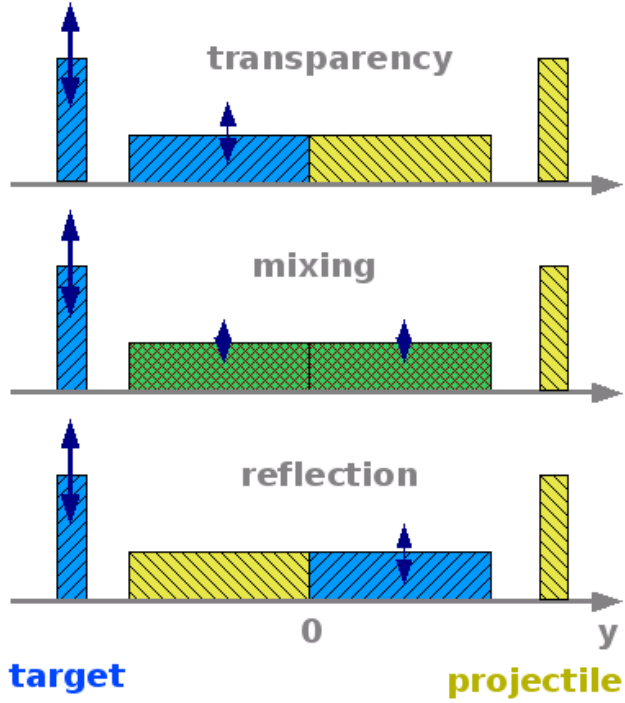,width=10cm} \caption{Sketch of the
rapidity distributions of the baryon number or the particle
production sources (horizontal rectangles) in nucleus-nucleus
collisions resulting from the transparency, mixing and reflection
models. The spectator nucleons are indicated by the vertical
rectangles. In the collisions with the fixed number of projectile
spectators only matter related to the target shows significant
fluctuations (vertical arrows). } \label{sketch}
\end{figure}

We are, however, far from a full understanding  of the A+A dynamics.
Many models based on  different assumptions compete 
with each other
and a consistent
description of all aspects of the data within a single model is
missing. The largest uncertainties concern the early stage
of collisions. It is unclear how initial nuclear flows of energy and
charges evolve. The majority of  the dynamical models (e.g. the
string-hadron transport approaches and the quark-gluon cascade
models \cite{hsd,urqmd,ampt}) predict or assume that the colliding
nuclear matter is transparent. The final longitudinal flows of the
hadron production sources or the net baryon number related to the
projectile and target follow the directions of the projectile and
target, respectively.~~~ We call this class of models transparency
(T-)models. Since the pioneering works of Fermi \cite{fermi} and
Landau \cite{landau} statistical and hydrodynamical approaches are
successfully used to describe high energy nuclear collisions. Many
models within this group, including the first Fermi formulation,
assume full equilibration of the matter at the early stage of
collisions. The initial projectile and target flows of energy and
charges  are mixed. The approaches which predict or suppose the full
mixing of the projectile and target flows we call the mixing
(M-)models. Let us note  that there are models which assume the
mixing of hadron production sources (inelastic energy) whereas the
transparency of baryon number flows, e.g. statistical model of the
early stage \cite{smes} and the three-fluid hydrodynamical model
\cite{3fluid}.
Finally, one may even speculate that the initial
flows are reflected in the collision process,
i.e. the flows of matter related to the target and the projectile
change their directions.
This class of models we call the reflection
(R-)models.
The sketch of the rapidity distributions resulting from
the T-, M- and R-models are shown in Fig.~\ref{sketch}.
The spectra related to the projectile and the target can be easily
distinguished in the figure because they are marked in color and
hatching the same way as the initial projectile and target nuclei.
In this paper we propose a  method to mark the matter related to
the projectile and the target in fluctuations (the MinF-method),
which allows to test experimentally different scenarios of the
collision process. Finally we apply the MinF-method to the NA49
experimental data on Pb+Pb collisions at 158$A$ GeV ($\sqrt{s_{NN}}$
= 17.2 GeV) \cite{maciek}.

\vspace{0.3cm} {\bf 2.} In each A+A collision only a part of all
2$A$ nucleons interact. These are called participant nucleons and
they are denoted as $N_P^{proj}$ and $N_P^{targ}$ for the projectile
and target nuclei, respectively. The nucleons which do not interact
are called the projectile and target spectators, $N_S^{proj} = A -
N_P^{proj}$ and $N_S^{targ} = A - N_P^{targ}$.
The fluctuations in high energy A+A collisions are dominated by a
trivial geometrical  variation of the impact parameter. However,
even for the fixed impact parameter the number of participants,
$N_P\equiv N_P^{proj}+N_P^{targ}$, fluctuates from event to event.
This is caused by the fluctuations of the initial states of the
colliding nuclei and the probabilistic character of an interaction
process. The fluctuations of $N_P$ usually  form a large and
uninteresting background. In order to minimize its contribution NA49
selected samples of collisions with  fixed numbers of projectile
participants. This selection is possible due to the measurement of
$N_S^{proj}$ in each individual collision by use of a calorimeter
which covers the projectile fragmentation domain. However, even in
the samples with $N_P^{proj} = const$ the number of target
participants fluctuates considerably. Hence, an asymmetry between
projectile and target participants is introduced, i.e. $N_P^{proj}$
is constant, whereas $N_P^{targ}$ fluctuates. This difference is
used in the  MinF-method to distinguish between the final state
flows related to the projectile and the target. Qualitatively, one
expects  large fluctuations of any extensive quantity (e.g. net
baryon number and multiplicity of hadron production sources) in the
domain related to the target and  small fluctuations in the
projectile region. When  both flows are mixed intermediate
fluctuations are predicted. The whole procedure is presented in a
graphical form in Fig.~\ref{sketch}. Clearly, the fluctuations
measured in the target momentum hemisphere are larger than those
measured in the projectile hemisphere in T-models. The opposite
relation is predicted for R-models, whereas for M-models the
fluctuations in the projectile and target hemispheres are the same.

This general qualitative idea is further on illustrated
by quantitative calculations performed within
several models ordered by an increasing
complexity.

\vspace{0.3cm} {\bf 3.} Let us begin with
considering  fluctuations of the net baryon number measured
in different regions of the participant domain in collisions of two
identical nuclei. These fluctuations are most closely related to the
fluctuations of the number of participant nucleons because of the baryon
number conservation. In the following the variance, $Var(x) \equiv
\langle x^2 \rangle - \langle x \rangle^2$, and the scaled variance,
$\omega_x \equiv Var(x)/\langle x \rangle$, where $x$ stands for a
given random variable and $\langle \cdots \rangle$ for
event-by-event averaging, will be used to quantify fluctuations. We
denote by $\omega_P^{targ} \equiv Var(N_P^{targ})/\langle N_P^{targ}
\rangle$ the scaled variance of the number of target participants
and by $\omega_B \equiv Var(B)/\langle B \rangle$ the scaled
variance of the net baryon number, $B$. In each event we subtract
the nucleon spectators when counting the number of baryons. The net
baryon number, $B\equiv N_B-N_{\overline{B}}$, equals then  the
number of participants $N_P=N_P^{targ}+N_P^{proj}$. At fixed
$N_P^{proj}$, the $N_P$ number fluctuates due to the fluctuations of
$N_P^{targ}$. 
The distribution in $N_P^{targ}$ can be characterized by its
mean value,
$\langle N_P^{targ}\rangle \simeq N_P^{proj}$, and a scaled
variance, $\omega_P^{targ}$. Thus, for the net baryon baryon number
$B$
one finds,
\eq{ \omega_B~=~\frac{Var(N_P)}{\langle N_P\rangle}~\simeq~
\frac{\langle \left(N_P^{targ}\right)^2\rangle~-~\langle
N_P^{targ}\rangle^2}{2\langle
N_P^{targ}\rangle}~=~\frac{1}{2}~\omega_P^{targ}~, \label{omegaB} }
for the fluctuations in the full phase space of participant
nucleons. A factor $1/2$ in the right hand side of
Eq.~(\ref{omegaB}) appears because only a half of the total number
of participants fluctuates. Let us introduce $\omega_B^p$ and
$\omega_B^t$, where the superscripts $p$ and $t$ mark quantities
measured in the projectile and target momentum hemispheres,
respectively.
 By assumption, the mixing of the projectile and target participants
is absent in T- and R-models. Therefore,
in T-models, the net baryon number in the projectile hemisphere
equals  $N_p^{proj}$ and does not fluctuate, i.e.
$\omega_B^{p}(T)=0$, whereas the net baryon number in the target
hemisphere equals  $N_p^{targ}$ and fluctuates with
$\omega_B^{t}(T)=\omega_P^{targ}$.
These relations are reversed in R-models.
%

%
%
%
%

We introduce now a random mixing of baryons between the projectile
and target hemispheres. Let $\alpha$ be a probability for
(projectile) target participant to be detected in the (target)
projectile hemisphere. It is easy to show that:
\eq{
\omega_B^t~=~(1~-~\alpha)^2~\omega_P^{targ}~+~2\alpha(1~-~\alpha)~,~~~~
\omega_B^p~=~\alpha^2~\omega_P^{targ}~+~2\alpha(1~-~\alpha)~.
\label{alpha}
}
 A (complete) mixing of the projectile and target participants
is assumed in M-models. Thus each participant nucleon with equal
probability, $\alpha=1/2$, can be found either in the target or in
the projectile hemispheres.
In M-models the fluctuations in both
hemispheres are identical. The limiting cases, $\alpha=0$ and
$\alpha=1$ of Eq.~(\ref{alpha}) correspond to T- and R-models,
respectively. In summary the scaled variances of the net baryon
number fluctuations in the projectile, $\omega_B^p$, and target,
$\omega_B^t$, hemispheres are:
\eq{
&\omega_B^{p}(T)~=~0~,~~~~~~
\omega_B^{t}(T)~=~\omega_P^{targ}~, \label{TB}\\
&\omega_B^{p}(M;~rr)~=~\omega_B^{t}(M;~rr)~=~\frac{1}{2}~+~
~\frac{1}{4}~ \omega_P^{targ}~,\label{MB}\\
&\omega_B^{p}(R)~=~\omega_P^{targ}~, ~~~~ \omega_B^{t}(R)~=~0~,
\label{RB} }
in the T- (\ref{TB}), M- (\ref{MB}) and R- (\ref{RB}) models of the
baryon number flow. 
When deriving Eq.~(\ref{MB}) we assumed that the baryons are 
distributed randomly in the rapidity space thus the abbreviation $rr$
in the left hand side of Eq.~(\ref{MB}) stands for $random$ $rapidities$.
This implies that even  for a fixed
number of $N_P^{targ}$, i.e. for $\omega_P^{targ}=0$, the baryon
number in the projectile and target hemispheres  fluctuates,
$\omega_B^{p}(M;~rr)~=~\omega_B^{t}(M;~rr)~=~1/2$. 

In  a mixing model in which  baryon rapidities do not
fluctuate from collision to collision, but their positions are
fixed (the $fixed$ $rapidity$, ({\it fr}), model)
the scaled variances in the projectile and target hemispheres read:
 \eq{
\omega_B^{p}(M;~fr)~=~\omega_B^{t}(M;~fr)~=~ \frac{1}{4}~
\omega_P^{targ}~.\label{MB-fr}}
Note that
the term $1/2$ of the right hand side of Eq.~(\ref{MB}) is
absent in (\ref{MB-fr}).

\begin{figure}[ht!]
\epsfig{file=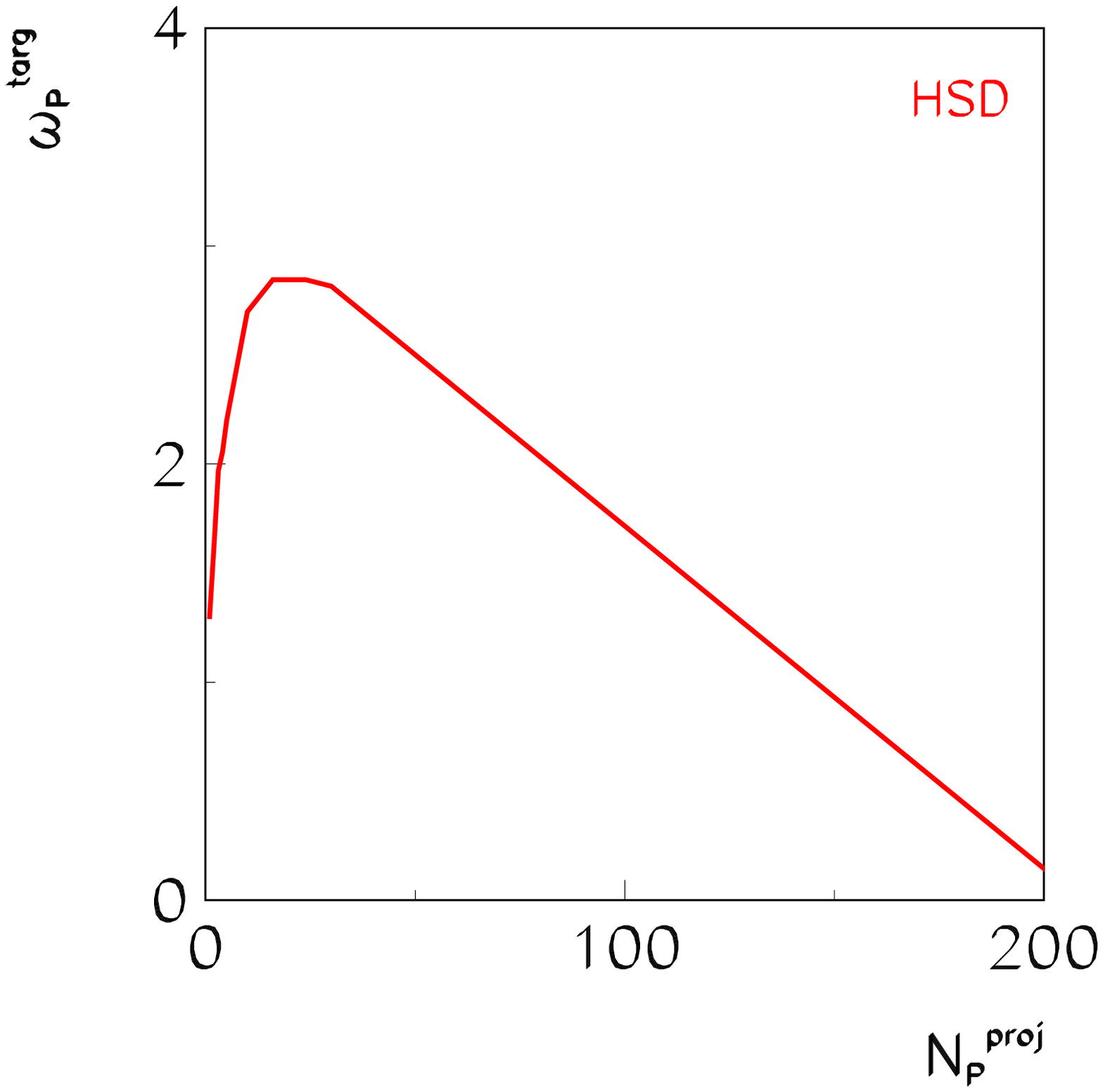,width=8cm}
 \epsfig{file=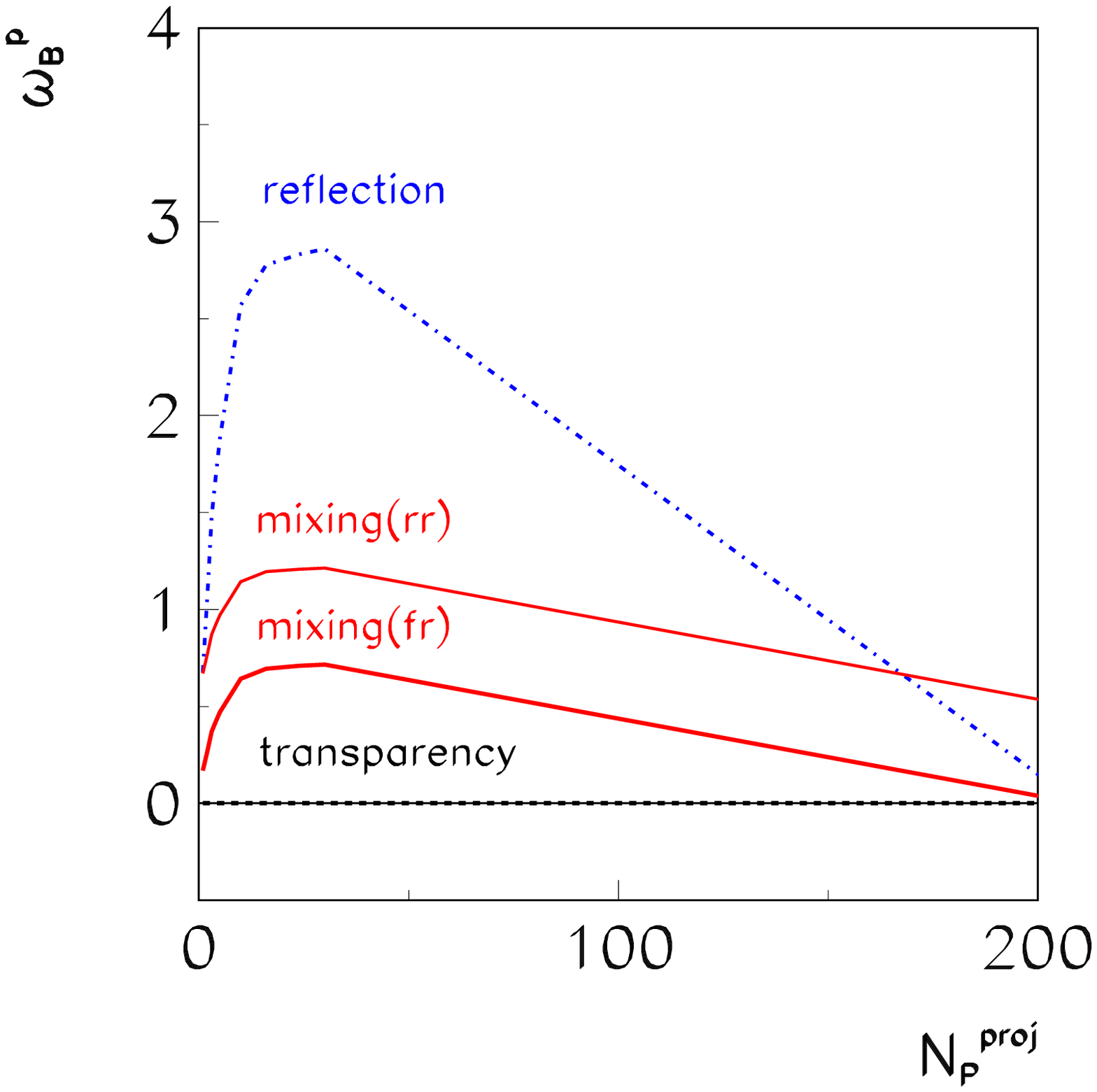,width=8cm}
 \caption{{\it Left}: The scaled variance
$\omega_P^{targ}$  for the fluctuations of target participants
$N_P^{targ}$ as a function of $N_P^{proj}$ calculated \cite{BHK}
within the HSD model. {\it Right}: The scaled variances
$\omega_B^{p}$ versus $N_P^{proj}$ obtained within T- (dashed line) 
and R-models (dashed-dotted line), Eqs.~(\ref{TB},~\ref{RB}). 
The upper solid line
shows predictions of the M-model with random rapidities of baryons 
(\ref{MB})
whereas the lower
solid line corresponds to the M-models with fixed rapidities of
baryons (\ref{MB-fr}). For
$\omega_B^t$ the predictions for T- and R-models
should be exchanged and
the lines for the M-models remain unchanged.   \label{omegaP-fig}}
\end{figure}

Eventually, for an estimate of the magnitude of the  expected
fluctuations in different type of models
we consider an example of Pb+Pb collisions at 158$A$ GeV.
The scaled variance of the number of target participants at the
fixed number of projectile participants 
(i.e. $\omega_P^{proj} = 0$)
can be calculated within the
string-hadronic models. The corresponding results \cite{BHK}
obtained for the HSD \cite{hsd} model are shown in
Fig.~\ref{omegaP-fig}, left.
 Using Eqs.~(\ref{TB}-\ref{MB}) 
and the dependence of 
$\omega_P^{targ}$ on $N_P^{proj}$ calculated within the HSD model
(Fig.~\ref{omegaP-fig} (left)), 
quantitative predictions concerning the baryon number fluctuations
for different  
models can be obtained. The resulting dependencies of the 
scaled variance of the baryon number in the projectile
hemisphere on $N_P^{proj}$ are shown in
Fig.~\ref{omegaP-fig} (right).
As expected large fluctuations are seen in R-models,
intermediate in M-models and there are no fluctuations in
T-models. In the M-model  the scaled variance increases
by 1/2 when baryon
positions in  rapidity are assumed to fluctuate.

\vspace{0.3cm} {\bf 4.} The T-, M- and R-models for the baryonic
flows give indeed very different predictions for $\omega_B^p$ and
$\omega_B^t$ for the samples of events with fixed values of
$N_P^{proj}$. However, they may be difficult to test experimentally
as an identification of protons and a measurement of neutrons in a
large acceptance  in a single event is difficult.
Measurements of charged particle multiplicity in a large acceptance
can be performed using  the existing detectors. 
In particular, the first results on multiplicity fluctuations 
of negatively charged hadrons, $N_-$,
as a function of $N_P^{proj}$  were recently obtained by
NA49~\cite{maciek} for
Pb+Pb collisions at 158$A$ GeV.  
Note that at the CERN SPS and
lower energies negatively charged hadrons are predominantly (more
than 90\%) $\pi^-$ mesons. In the following we consider T-, M- and
R- scenarios within several approaches to particle production in
high energy nuclear collisions. We suppose that a part of the
initial projectile and target energy, the inelastic energy, is
converted into hadron sources. Further on, the numbers of projectile
and target related sources are taken to be proportional to the
number of projectile and target participant nucleons, respectively.
The physical meaning of a particle source depends upon the model
under consideration, examples are  wounded nucleons (see
\cite{wnm,wnm1}), strings and  resonances (see \cite{hsd,urqmd}),
and volume cells of the expanding matter at the freeze-out in the
hydrodynamical models. For the independent sources one finds regarding  the
scaled variance of $i$-th particle species (see e.g. \cite{source}):
\eq{
\omega_i~=~\omega_i^*~+~\langle n_i^* \rangle ~\Omega_*~,
\label{omega-i} }
where $\omega_i^*$  denotes the scaled variance for $i$-th hadron
species (e.g., $i$ may correspond to $h^-$) from a single source,
$\langle n_i^* \rangle $ is the average multiplicity from a single
source, and $\Omega_*$ is the scaled variance for the fluctuation of
the number of sources. Assuming that the number of hadron sources
is proportional to the number of participating nucleons,
$N_*=const\cdot N_P$,
one gets:
\eq{
\langle n_i^*\rangle ~\Omega_* ~\equiv~\frac{\langle
N_i\rangle}{\langle N_*\rangle} ~\Omega_*~=~\frac{\langle
N_i\rangle}{\langle N_P\rangle} ~\omega_P~\equiv~
\overline{n}_i~\omega_P~,
\label{ni} }
where $\overline{n}_i$ is the average  multiplicity 
of the $i$-th species per
participating nucleon. Thus the scaled variance~(\ref{omega-i}) of
the particle number multiplicity   in the full phase space is:
\eq{
\omega_i~=~\omega_i^*~+~\overline{n}_i
~\frac{1}{2}~\omega_P^{targ}~.
\label{omega1-i} }
Consequently, the scaled variances of the $i$-th hadron multiplicity
distribution in T-, M- and R-models read:
 \eq{
&\omega_i^{p}(T)~=~\omega_i^{*}~,~~~~
\omega_i^{t}(T)~=~\omega_i^*~+~
\overline{n}_i~\omega_P^{targ}~, \label{T-}\\
&\omega_i^{p}(M;~rr)~=~\omega_i^{t}(M;~rr)~=~ \omega_i^*~+
~\overline{n}_i ~\left(~\frac{1}{2}~+~\frac{1}{4}~
\omega_P^{targ}~\right)~,\label{M-}\\
&\omega_i^{p}(M;~fr)~=~\omega_i^{t}(M;~fr)~=~ \omega_i^*~+
~\overline{n}_i ~\frac{1}{4}~
\omega_P^{targ}~,\label{Mfr-}\\
& \omega_i^{p}(R)~=~\omega_i^*~+~ \overline{n}_i~
\omega_P^{targ}~,~~~~ \omega_i^{t}(R)~=~\omega_i^*~. \label{R-} }
%
%
Again two different versions of mixing with {\it random rapidities}
(\ref{M-}) and {\it fixed rapidities} (\ref{Mfr-}) of the source
positions are possible. As an example of M-models with fixed
rapidities of the sources let us consider a model which assumes a
global equilibration of the matter at the early stage of collisions
followed by a hydrodynamical expansion and freeze-out. In this case
particle production sources can be identified with the volume cells
of the expanding matter at the freeze-out. They can be treated as
uncorrelated provided the effects of global energy-momentum
conservation laws can be neglected. Due to assumed global
equilibration of the projectile and target flows the fluctuations in
the projectile and target hemispheres are identical. The model
belongs to the class of M-models.  In this model there is one to one
correspondence between space-time positions and rapidities of the
hydrodynamic cells. Thus, the source rapidities do not fluctuate and
the scaled variances of hadrons in the projectile and target
hemispheres have the form (\ref{Mfr-}).

 Note that Eqs.~(\ref{T-}) and (\ref{R-}) are
strictly valid provided that a source produces particles only in its
hemisphere. Due of the finite width of the rapidity distribution
resulting from the decay of a single source this condition is
expected to be violated at least close to midrapidity. Thus, in
order to be able to neglect the cross-talk of particles between the
projectile and target hemispheres the width of the rapidity
distribution of a single source, $\Delta y_*$, should be much
smaller than the total width of the rapidity
distribution.

Some comments concerning Eqs.~(\ref{T-}-\ref{R-}) are appropriate.
There is a general similarity of the expressions for produced
particles  and the corresponding expressions for baryons,
Eqs.~(\ref{TB}-\ref{MB-fr}). There are, however, two important
differences. A single source produces particles in a probabilistic
way  with an average multiplicity  $\langle n_i^*\rangle$ and a
scaled variance $\omega_i^*$. Consequently, it leads to an
additional term, $\omega_i^*$, in all expressions for
$\omega_i^{p,t}$,
and an additional factor, $\overline{n}_i$, appears in terms related
to the fluctuations of the number of sources. Following
Eq.~(\ref{ni}) the source number fluctuations can be substituted by
$\omega_P^{targ}$, and an average  multiplicity, $\langle
n_i^*\rangle$, of a single source can be then transformed into an
average multiplicity per participating nucleon, $\overline{n}_i$.
The term, 1/2, in the r.h.s.  of Eq.~(\ref{M-}), as that in
Eq.~(\ref{MB}), is due to the random rapidity positions of the
sources in M-models.
%
In the hydrodynamical model
particle production sources can be identified with the volume cells
of the expanding matter at the freeze-out.
The source rapidities do not fluctuate and Eq.~(\ref{M-}) is
transformed then into Eq.~(\ref{Mfr-}).

\begin{figure}[h!]
\epsfig{file=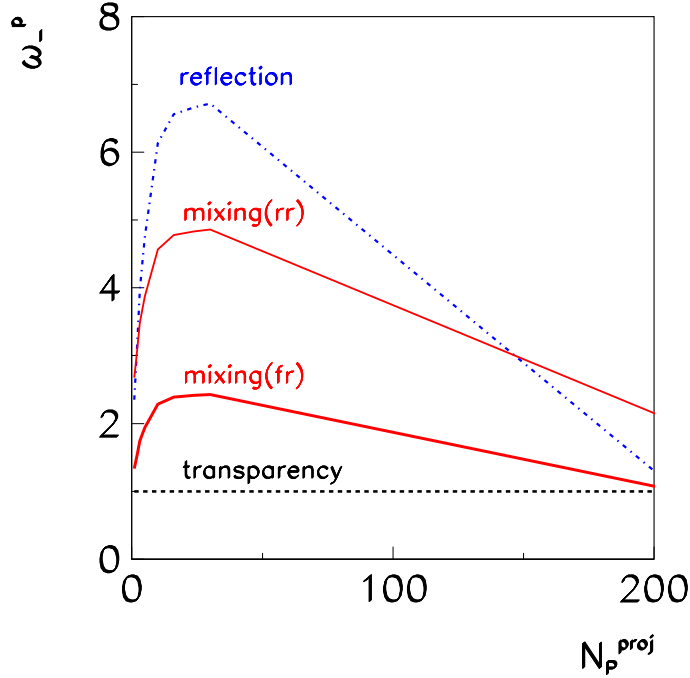,width=10cm} \caption{The dependence of the
scaled variance of negatively charged particle multiplicity in Pb+Pb
collisions at 158$A$ GeV on the number of projectile participants,
$N_P^{proj}$, in the projectile hemisphere.  The predictions for T-
(\ref{pTtR}) and R- (\ref{tRpT}) models are shown by dashed and
dashed-dotted lines, respectively. The upper solid line, mixing(rr),
corresponds to the M-models with  random rapidity positions of
the sources (\ref{pMtM}). The lower solid line, denoted as
mixing(fr), corresponds to the M-models with fixed rapidity
positions of the sources (\ref{pMtM-fr}). For the target hemisphere
the lines of T- and R-models should be interchanged, whereas the
lines of M-models remain unchanged. We take $\omega_P^{targ}$ from
Fig.~\ref{omegaP-fig} (left) for all types of models. }
\label{omegamin-fig}
\end{figure}

We turn now to a discussion of  multiplicity fluctuations
of negatively charged hadrons in Pb+Pb
collisions at 158$A$~GeV. The value of
$\langle N_- \rangle/\langle N_P \rangle \equiv \overline{n}_-\simeq
2$ was measured for the studied reactions \cite{peter}. For
simplicity we assume $\omega_-^*\simeq 1$, this is valid for the
poissonian negatively charged particle multiplicity distribution from a
single source. Note, in p+p interactions at SPS energies and in 
the limited acceptance of NA49 the measured distribution is in
fact close to the Poisson one \cite{maciek}.
This gives:
 \eq{
&\omega_{-}^{p}(T)~=~ \omega_{-}^{t}(R)~\simeq~1~,\label{pTtR} \\
&\omega_{-}^{p}(M;~rr)~=~\omega_-^t(M;~rr)~\simeq~ 2~+~\frac{1}{2}~
\omega_P^{targ}~,\label{pMtM}\\
&\omega_{-}^{p}(M;~fr)~=~\omega_-^t(M;~fr)~\simeq~ 1~+~\frac{1}{2}~
\omega_P^{targ}
\label{pMtM-fr} \\
&\omega_{-}^{t}(T)~=~
\omega_{-}^{p}(R)~\simeq~1~+~2~\omega_P^{targ}~.\label{tRpT}
}
The dependence of $\omega_-^p$ on $N_P^{proj}$ from
Eqs.(\ref{pTtR}-\ref{tRpT}) for T-, M- and R-models is presented in
Fig.~\ref{omegamin-fig}. .

%
%

%
%
%

\vspace{0.3cm} {\bf 5.}
In a recent analysis \cite{wnm1} of d+Au interactions at
$\sqrt{s_{NN}} = 200$~GeV 
\cite{Back:2004mr}
within the wounded nucleon model (WNM)
\cite{wnm} it was found that the wounded nucleon sources emit
particles in a very broad rapidity interval which results in the mixing
of particles from the target and projectile sources. In the
following we consider predictions of this model with respect to
multiplicity fluctuations.

The WNM assumes that $i$-th particle rapidity distribution in A+A
collisions is presented as
\eq{
\frac{dN_i}{dy}~=~N_P^{targ}~F_i^{t}(y)~+~N_P^{proj}~F_i^{p}(y)~,
\label{wnm-y-spectr} }
where $F_i^{t}(y)$ and $F_i^{p}(y)$ are the contributions from a
single wounded nucleon (identified with a particle source)
of the target and
projectile, respectively. The model also requires
\eq{ F_i^p(y)~=~F_i^t(-y)~
\label{Fip-Fit} }
in the center of mass system of the collision. The mean number of
particles in the rapidity interval $\Delta y$
for collisions with $N_P^{proj}$ and   $N_P^{targ}$
is given by
\eq{ N_i(\Delta y)~=~N_P^{targ}~\int_{\Delta y} dy~
F_i^{t}(y)~+~N_P^{proj}~\int_{\Delta y} dy~ F_i^{p}(y)~.
\label{Ni-Delta-y} }
For interaction of identical heavy ions
$\langle N_P^{targ}\rangle \simeq
N_P^{proj}$, and then  Eq.~(\ref{Ni-Delta-y}) yields:
\eq{ \langle N_i(\Delta y)\rangle ~=~N_P^{proj}~\int_{\Delta y} dy~
\left[F_i^{t}(y)~+~ F_i^{p}(y)\right]~.
\label{Ni-Delta-y-av} }
Let us consider now fluctuations of $N_i(\Delta y)$ at a fixed
$N_P^{proj}$.
A contribution to the scaled variance of
$N_i(\Delta y)$ (\ref{Ni-Delta-y}) due to the
fluctuations of $N_P^{targ}$ reads:
\eq{
\frac{Var[N_i(\Delta y)]}{\langle N_i(\Delta y)\rangle}
~=~\frac{\left[\int_{\Delta y} dy~ F_i^{t}(y)\right]^2}{\int_{\Delta
y} dy~ \left[F_i^{t}(y)~+~
F_i^{p}(y)\right]}~\omega_P^{targ}~\equiv~  n_i^t(\Delta
y)~\alpha^t_i(\Delta y)~\omega_P^{targ}~,\label{omegai-delta-y}
}
where
\eq{n_i^t(\Delta y)~\equiv~\int_{\Delta y} dy~ F_i^{t}(y)~,~~~~
\alpha^t_i(\Delta y) ~\equiv~ \frac{\int_{\Delta y} dy~
F_i^{t}(y)}{\int_{\Delta y} dy~ \left[F_i^{t}(y)~+~
F_i^{p}(y)\right]}~. \label{ni-R}
}
As previously, for simplicity we assume that 
a single source emits particles according to
the Poisson distribution,
$\omega_i^*(\Delta y)=1$. 
This leads to a general expression on  the scaled
variance for a particle of $i$-th type:
\eq{
\omega_i(\Delta y)~=~
1~+ n_i^t(\Delta y)~\alpha^t_i(\Delta
y)~\omega_P^{targ}~.\label{omegai1-delta-y}
}
The parameter $\alpha^t_i$ quantifies the amount of
mixing of the projectile and target contributions and
can vary between 0 and 1. 
For full acceptance, $\Delta y =[-Y_{max},Y_{max}]$,
Eq.~(\ref{omegai1-delta-y}) transforms  to
Eq.~(\ref{omega1-i}).

The T-, M- and R- limits of the WNM can be formulated in terms of
the distribution functions of the single nucleon,
\eq{
& F_i^p(y;T)~=~T_i(y)~\theta(y)~,~~~~ F_i^t(y;T)~=~T_i(-y)~\theta(-y)~, \label{FT}\\
& F_i^p(y;M)~=~F_i^t(y;M)~=~M_i(y)~,\label{FM}\\
& F_i^p(y;R)~=~R_i(-y)~\theta(-y)~,~~~~
F_i^t~=~R_i(y)~\theta(y)~.\label{FR}
}

The scaled variances in the projectile and target hemispheres,
$\omega_i^p\equiv \omega_i(y\ge 0)$ and $\omega_i^t\equiv
\omega_i(y\le 0)$, can be found using Eq.~(\ref{omegai1-delta-y}).
It follows that in T- and R-models, the $\omega_i^{p,t}$ coincide
with those given by Eq.~(\ref{T-}) and Eq.~(\ref{R-}),
respectively. For the M-models this gives the following result:
\eq{
\omega_i^{p}(M;~WNM)~=~\omega_i^{t}(M;~WNM)~=~
1~+ ~\overline{n}_i ~\frac{1}{4}~ \omega_P^{targ}~,\label{M-WNM}
}
which is identical to Eq.~(\ref{Mfr-}). In the WNM all projectile
(target) sources are assumed to be identical and their positions are
the same and fixed. Therefore, similar to the hydrodynamical model,
the term, 1/2, in the r.h.s. of Eq.~(\ref{M-}) is absent in
Eq.~(\ref{M-WNM}). This is the M-model with the {\it fixed rapidity}
positions of the sources. The mixing in the considered version of
WNM results from a broad distribution of particles produced by a
single source. A complete mixing in the WNM means according to
Eq.~(\ref{FM}) that projectile and target source functions become
identical.

 \vspace{0.3cm}
 {\bf 6.}
Let us consider fluctuations in  limited phase-space domains in
which only  fractions, $q^{p,t}$, of all particles in the projectile
or target hemispheres are accepted. Then the scaled variances in the
acceptance,
 $\omega^p_{acc}$ and $\omega^t_{acc}$, will be different
from the $\omega^p$ and $\omega^t$.
We start with the scaled variances $\omega_B^p$ and $\omega_B^t$
of the net baryon number fluctuations. Assuming that inside the
projectile and target hemispheres the baryon rapidities are not
correlated one gets (see e.g.,\cite{source,ce1}):
\eq{
\omega^{p,t}_{B,acc}~ = ~1~-~q^{p,t}~+~q^{p,t}\cdot\omega_B^{p,t}~.
\label{acc-pt} }

It can be  shown that the scaled variance of the produced particles
in the limited momentum acceptance within the M-model with fixed
source rapidities (\ref{Mfr-}) reads:
\eq{
\omega_{i,acc}(M,~fr) = 1 + \frac {1} {2} \frac {\langle N_i
\rangle_{acc}} {\langle N_P \rangle}  \omega_P^{targ}~, \label{hyd}
}
where $\langle N_i \rangle_{acc}$ is a mean  multiplicity of
a particle of $i$-th type 
in the acceptance. The formula above assumes that the
produced  particles are uncorrelated in the momentum space, i.e. it
neglects effects of motional conservation laws and resonance decays.
The scaled variance in a limited rapidity acceptance, $\Delta y$,
within WNM can be directly obtained from
Eq.~(\ref{omegai1-delta-y}). It coincides with that of
Eq.~(\ref{hyd}).

Let us consider now as an example the
NA49 acceptance, which is located in the projectile hemisphere about
one and half rapidity units from mid-rapidity, $\Delta y = [1.1;~
2.6]$ in the c.m.s. The acceptance probability was measured to be
$q^p \simeq 0.4$~\cite{maciek,kasia} (i.e. about 40\% of negatively
charged particles in the projectile hemisphere are accepted).
In the limiting case of the fixed rapidities of the sources, this is
assumed to be valid for both the hydrodynamical and WN models,
one finds:
\eq{
&\omega_{-,acc}^{p}(T)~\simeq~1, \label{accT}\\
& \omega_{-,acc}^{p}(M;~fr)
~\simeq~
1~+~0.2~\omega_P^{targ}~,\label{accM} \\
& \omega_{-,acc}^{p}(R;~fr)~\simeq~
1~+~ 0.8~\omega_P^{targ}~.\label{accR}
}
The relations, $n_-^t(\Delta y; T)=0;~~n_-^t(\Delta y; M)~=~0.5~
q^p~ \overline{n}_-;~~n_-^t(\Delta y; R)~=~ q^p~ \overline{n}_-$,
with $q^p~\simeq~0.4$ and $\overline{n}_-~\simeq~2$ have been used in
Eqs.~(\ref{accT}-\ref{accR}).
Note that in the limit $q^p\rightarrow 0$ one finds
$\omega_{-,acc}^{p}\simeq 1$ for all type of models.

\begin{figure}[ht!]
\epsfig{file=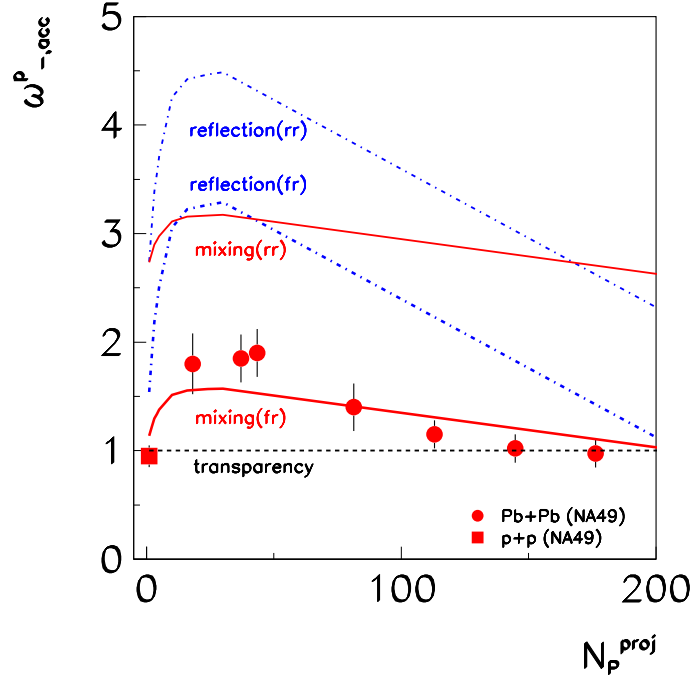,width=10cm} \caption{The dependence of
the scaled variance of negatively charged particle multiplicity in
Pb+Pb collisions at 158$A$ GeV on the number of projectile
participants, $N_P^{proj}$, in the NA49 acceptance located in the
projectile hemisphere. The Pb+Pb data \cite{maciek} are indicated by
filled circles. For a comparison the  result for p+p interactions at
158 GeV \cite{maciek,kasia} is shown by the filled square. The
displayed errors correspond to the sum of systematic and statistical
uncertainties. The dashed line shows a dependence predicted within
T-models (Eqs.~(\ref{accT}) and (\ref{acc-T})), the solid lines
correspond to the mixing(fr) models (lower line, Eq.~(\ref{accM})),
and the mixing(rr) models (upper line, Eq.~(\ref{acc-M})). The lower
dashed-dotted line corresponds to the reflection(fr) models
(\ref{accR})), whereas the reflection(rr) models (\ref{acc-R}) are
indicated by the upper dashed-dotted line.
 \label{data}}
\end{figure}

The predictions given by Eqs.~(\ref{accT}-\ref{accR}) 
are shown in Fig.~\ref{data}. One
may be surprised that different models lead to the same results for
most central collisions. This is because  $\omega_P^{targ}$ goes
to zero at $N_P^{proj}\simeq A$, as it follows from
Fig.~\ref{omegaP-fig} (left). 
The predictions of the
T-, M- and R-models differ because of  their different response
on the $N_P^{targ}$ fluctuations. These fluctuations
become small in the most central events. Therefore, the best way to
study the mixing-transparency effects is the analysis of the {\it
centrality dependence} of the particle number fluctuations in the
projectile and target hemispheres.

\vspace{0.2cm} We now discuss an effect of a limited acceptance for
the approaches with randomly fluctuating source rapidities. In this
case the scaled variances in the projectile and target hemispheres
are given by Eqs.~(\ref{T-}-\ref{R-}), provided a width of a
rapidity spectrum of particles emitted from a single source is
narrow.
  In a general case, when a rapidity width of the source particles
and an acceptance window are comparable in size, it is  difficult
to make analytical estimates. The problem can be solved in the limit
of very narrow sources (a source width, $\Delta y_*$, is much
smaller than the experimental acceptance interval, $\Delta y$). The
hadrons created by "narrow" sources have  correlated rapidities, but
Eq.~(\ref{acc-pt}) can be used for the scaled variances of the
number of sources assuming that the source rapidities in the
projectile hemisphere are not correlated. In this case our
Eqs.~(\ref{T-}-\ref{R-}) yield:
\eq{
&\omega_{-,acc}^{p}(T;~rr)~\simeq~1, \label{acc-T}\\
& \omega_{-,acc}^{p}(M;~rr)~\simeq~ 1~+~ \overline{n}_-~ \left[
1~-~q^t~+~
q^t\cdot\left(\frac{1}{2}~+~\frac{1}{4}~\omega_P^{targ}\right)\right]
~\simeq~2.6~+~0.2~\omega_P^{targ}~,\label{acc-M} \\
& \omega_{-,acc}^{p}(R;~rr)~\simeq~1~+~ \overline{n}_-~ \left[
1~-~q^t~+~ q^t\cdot~\omega_P^{targ}\right]~\simeq~2.2~+~
0.8~\omega_P^{targ}~.\label{acc-R}
}
As before, we use $q^p~\simeq~0.4$ and $\overline{n}_-~\simeq~2$ in
Eqs.~(\ref{acc-T}-\ref{acc-R}). The corresponding curves are plotted
in Fig.~\ref{data}.

 The experimental points for  Pb+Pb collisions
at 158$A$ GeV clearly exclude transparency and reflection approaches
discussed here. The mixing model with  random fluctuations of a
source rapidity and a narrow width also strongly disagree with the
data. A reasonable agreement is observed only for the
mixing-hydrodynamical and WNM models. 
We remind that a large degree of mixing was found previously
in the analysis of the pseudo-rapidity spectra of charged
hadrons produced  in d+Au interactions at $\sqrt{s_{NN}} =$~200~GeV  
\cite{Back:2004mr} within WNM \cite{wnm1}.

Note however that the WNM model, used
here as a simple example to illustrate the MinF method, can not
reproduce many observables  connected to collective behavior of
matter created in high energy nuclear collisions, like radial and
anisotropic flows and the strangeness enhancement. On the other
hand, these effects are at least qualitatively described by
statistical and hydrodynamical approaches.

\vspace{0.3cm} {\bf 7.} At the end several comments are appropriate.
We considered three limiting behaviors of nuclear flows:
transparency, mixing and reflection. In general, all intermediate
cases are possible and they can be characterized by an
additional parameter.
Eq.~(\ref{alpha}) introduces a mixing parameter $\alpha$ for the
net baryon number with limiting cases $\alpha(T)=0$,
$\alpha(M)=1/2$, and $\alpha(R)=1$ for T-, M- and R-models,
respectively.  Within WNM  a parameter, $\alpha_i^t(\Delta y)$
(\ref{ni-R}), defined for each particle species, $i$, and for  each
rapidity interval, $\Delta y$, was suggested. The limiting cases are
again: $\alpha_i^t(\Delta y;T)=0,~ \alpha_i^t(\Delta y;M)=0.5$, and
$\alpha_i^t(\Delta y;R)=1$. The values of the mixing parameter can
be extracted by fitting the experimental data.

The fluctuations of the participant number  lead to the fluctuations
of the center of mass rapidity ($\Delta y\simeq -1/2
\log(N_P^{targ}/N_P^{proj})$. This alone may result in additional
multiplicity fluctuations. We estimated that for the NA49 data
discussed above the corresponding increase of the scaled variance is
smaller than 5\%.

The MinF-method can be used independently of the
degrees of freedom relevant at an early stage of collisions (e.g.
hadrons at a low collision energy or quark and gluons at a high
energy). This is because the concepts of the spectators and  the
participants as well as hadron multiplicity fluctuations are valid
at all relativistic energies and for all collision scenarios. In the
case of collisions of non-identical nuclei (different baryon numbers
and/or electric charge to baryon ratios) one can trace flows of the
conserved charges by looking at their inclusive final state
distributions (see e.g. \cite{fopi}). An interesting information can
be extracted from collisions of two nuclei with different atomic
numbers (see \cite{wnm1}).

 In the case of identical nuclei only the
MinF method can be used. It gives a unique possibility to
investigate the flows of both the net baryon number and particle
production sources.

\vspace{0.3cm} {\bf 8.} In summary, a method which allows to find
out what  happens with the initial flows in high energy
nucleus-nucleus collisions was proposed. First, the projectile and
target initial flows are marked in fluctuations (the MinF-method) in
the number of colliding nucleons. This can be achieved by  a
selection of collisions with a fixed number of projectile
participants but a fluctuating number of target participants. This
case is considered in details in the present study. Other selections
are also possible. Secondly, the projectile and target related
matter in the final state of collisions are distinguished by an
analysis of fluctuations of extensive quantities. We apply this
method to the NA49 data on multiplicity fluctuations of negatively
charged hadrons produced in Pb+Pb collisions at
158$A$~GeV~\cite{maciek}. The results
are consistent with the model which assumes
a significant degree
of mixing of the projectile and target flows at the early
stage of collisions followed by the hydrodynamical
expansion and freeze-out.

\vspace{0.3cm}
 We would like to thank V.V.~Begun, A.~Bialas, E.L.~
Bratkovskaya, S.~Haussler, Yu.B.~Ivanov, V.P.~ Konchakovski,
B.~Lungwitz, I.N.~Mishustin, St.~Mr\'owczy\'nski, M.~Rybczy\'nski,
H.~St\"ocker, and Z.~ W{\l}odarczyk for numerous discussions.
Moreover, we are grateful to Marysia Ga\'zdzicka for help in the
preparation of the manuscript. The work was supported in part by US
Civilian Research and Development Foundation (CRDF) Cooperative
Grants Program, Project Agreement UKP1-2613-KV-04 and Virtual
Institute on Strongly Interacting Matter (VI-146) of Helmholtz
Association, Germany.

\end{document}